\documentclass[%
 aip,
 amsmath,amssymb,
 reprint,%
]{revtex4-1}

\usepackage{graphicx}
\usepackage{dcolumn}
\usepackage{bm}

\usepackage[utf8]{inputenc}
\usepackage[T1]{fontenc}
\usepackage{mathptmx}
\usepackage{etoolbox}
\usepackage{color}

\makeatletter
\def\@email#1#2{%
 \endgroup
 \patchcmd{\titleblock@produce}
  {\frontmatter@RRAPformat}
  {\frontmatter@RRAPformat{\produce@RRAP{*#1\href{mailto:#2}{#2}}}\frontmatter@RRAPformat}
  {}{}
}%
\makeatother
\begin{document}

\preprint{AIP/123-QED}

\title[Flexibility of the factorized form of the unitary coupled cluster ansatz]{Flexibility of the factorized form of the unitary coupled cluster ansatz}
\author{Jia Chen}
 \affiliation{Department of Physics, University of Florida, Gainesville, FL 32611, USA}
\affiliation{Quantum Theory Project,  University of Florida, Gainesville, FL 32611, USA}
\author{Hai-Ping Cheng}%
\affiliation{Department of Physics, University of Florida, Gainesville, FL 32611, USA}
\affiliation{Quantum Theory Project,  University of Florida, Gainesville, FL 32611, USA}

\author{J. K. Freericks}
 \affiliation{Department of Physics, Georgetown University, 37th St. and O St., NW, Washington, DC 20057, USA}
\email{James.Freericks@georgetown.edu}
\email{hping@ufl.edu}
\email{jiachen@ufl.edu}

\date{\today}

\begin{abstract}
The factorized form of the unitary coupled cluster ansatz is a popular state preparation ansatz for  electronic structure calculations of molecules on quantum computers. It often is viewed as an approximation (based on the Trotter product formula) for the conventional unitary coupled cluster operator. In this work, we show that the factorized form is quite flexible, allowing one to range from conventional configuration interaction, to conventional unitary coupled cluster, to efficient approximations that lie in between these two. The variational minimization of the energy often allows simpler factorized unitary coupled cluster approximations to achieve high accuracy, even if they do not accurately approximate the Trotter product formula. This is similar to how quantum approximate optimization algorithms can achieve high accuracy with a small number of levels.
\end{abstract}

\maketitle

\section{\label{sec:introduction}Introduction:}

The electronic structure of molecules is viewed as one of the most promising applications of quantum computing to the field of chemistry~\cite{aspuru-guzik-review}. Within electronic structure, there are two promising pathways. The first is via quantum phase estimation, which performs time evolution on an initial state~\cite{kitaev} and extracts the energy by measuring a complex phase. It does so using controlled time-evolution to allow for Fourier signal processing of the accumulated phases---at the end of the algorithm, it collapses to an eigenstate and the accumulated phase tells us the eigenvalue. If the initial state is a superposition of states with a high amplitude for the ground state, the method will eventually determine the ground-state eigenvalue and will also prepare the ground state. This method results in extremely deep circuits (due to the controlled time evolution), and so it is not practical on computers available in the near term. The second is via the variational quantum eigensolver (VQE)~\cite{vqe}. This approach uses an ansatz to approximately prepare a ground-state wavefunction, measures the energy (using a break-up of the Hamiltonian into a sum over unitary operators that can each be directly measured), and then uses a classical computer to optimize the parameters in the wavefunction, repeatedly looping through this algorithm to complete the variational calculation. VQE has many different varieties, based on different strategies for preparing the target state and determining how to update it. Some examples include the ADAPT method~\cite{adapt}, which chooses the next operator to use in the state-preparation ansatz from an operator pool, hardware-efficient approaches~\cite{qcc}, which simply entangle the wavefunction (rather than applying fermionic excitations to a reference state) and then optimize the entanglement for the best energy, and methods that enlarge the wavefunction scope by including additional variational terms in a virtual fashion~\cite{bert}.

In all of the variational methods, we need to apply operators to some reference state, to prepare the state for the measurement phase. In this work, we focus on methods that use fermionic excitations. Coupled cluster is the gold standard for electronic structure calculations of weakly correlated molecules. In a conventional coupled-cluster calculation, we create a state by applying excitations to a reference state, in the form $|\psi\rangle=e^{\hat{T}}|\psi_0\rangle$, where $|\psi_0\rangle$ is the initial reference state (which we will take to be the Hartree-Fock state) and the excitation operator is a sum of excitation operators of different orders $\hat{T}=\hat{T}_1+\hat{T}_2+\hat{T}_3+\cdots$. Each operator of a given order includes all possible excitations from real orbitals present in the reference state to virtual orbitals used in the basis set included in the calculation (with amplitudes chosen to optimize the energy). In many cases, a number of amplitudes for particular excitation operators are zero, implying they are not included in the ansatz. For example, the singles and doubles excitations can be written schematically as
\begin{align}
    \hat{T} &=\hat{T}_1+\hat{T_2}+\cdots , \\
    &= \sum_i^{occ}\sum_a^{vir} \theta_i^a \hat{a}_a^{\dagger} \hat{a}_i^{\phantom{\dagger}} +  \sum_{ij}^{occ}\sum_{ab}^{vir} \theta_{ij}^{ab} \hat{a}_{a}^{\dagger}\hat{a}_{b}^{\dagger} \hat{a}_{j}^{\phantom{\dagger}}\hat{a}_{i}^{\phantom{\dagger}} + \cdots ~.
\end{align}
Here, we use letters from the beginning of the alphabet $a$, $b$, $c$, $\cdots$ to represent the virtual (unoccupied) spin orbitals available in the basis set, and letters from the middle of the alphabet $i$, $j$, $k$, $\cdots$ to represent the occupied (real) spin orbitals that appear in the reference state. The operators $\hat{a}_r$ ($\hat{a}_r^\dagger$) destroy (create) an electron in the spin-orbital labelled by $r$ and satisfy the canonical anticommutation relations. The singles amplitudes are denoted by $\theta_i^a$, the doubles amplitudes by $\theta_{ij}^{ab}$, and so on---these amplitudes represent real numbers, which can be equal to 0.

In conventional coupled cluster, we do not actually form the variational wavefunction. Instead, we perform a similarity transformation on the Hamiltonian, $\hat{\mathcal{H}}\to e^{-\hat{T}}\hat{\mathcal{H}}e^{\hat{T}}$, and then force the overlaps of all elemental excitations with the transformed Hamiltonian acting on the reference state to vanish; this effectively zeroes out the off-diagonal elements of the Hartree-Fock row of the transformed Hamiltonian. This then produces the so-called amplitude equations. The similarity transformation can be carried out exactly, because the Hadamard lemma $e^{\hat{A}}\hat{B}e^{-\hat{A}}=\hat{B}+[\hat{A},\hat{B}]+\frac{1}{2}[\hat{A},[\hat{A},\hat{B}]]+\cdots$ involving a sum of terms with increasingly nested commutators, truncates after the fourth-order term because the Hamiltonian only has single and two-body operators in it. Note that this standard form of coupled cluster is no longer a variational calculation.

Unitary coupled cluster (UCC) is usually carried out in a variational fashion, which makes it much less efficient than conventional coupled cluster. In UCC, we form the variational wavefunction via
\begin{equation}
    |\psi_{UCC}\rangle=e^{\hat{T}-\hat{T}^\dagger}|\psi_0\rangle. 
\end{equation}
In this case, the Hadamard lemma does not generically truncate, so one is forced to work with the wavefunction directly. This comes at a huge computational cost, making UCC inefficient on classical computers. But, on quantum computers, it is feasible, if one can prepare the UCC operator in an efficient way and apply it to the reference state; especially so, since conventional coupled cluster cannot be carried out on a quantum computer. In general, this is difficult for the general form of the ansatz. This is because we do not know how to write general quantum circuits for sums of operators in an exponential (however, this may be changing~\cite{babbush_new}). Instead, we use a Trotter product formula to break the conventional UCC approximation up into a product of factors for which quantum circuits are known. This has us rewrite the UCC ansatz in a Trotter product form as 
\begin{align}
|\psi_{UCC}\rangle&=\lim_{N\to \infty}\left (\prod_{ia}e^{\frac{1}{N}\theta_i^a(\hat{a}_a^\dagger\hat{a}_i^{\phantom{\dagger}}-\hat{a}_i^\dagger\hat{a}_a^{\phantom{\dagger}})}\right .\nonumber\\&\times\left .\prod_{ijab}e^{\frac{1}{N}\theta_{ij}^{ab}( \hat{a}_{a}^{\dagger}\hat{a}_{b}^{\dagger} \hat{a}_{j}^{\phantom{\dagger}}\hat{a}_{i}^{\phantom{\dagger}}-\hat{a}_{i}^{\dagger}\hat{a}_{j}^{\dagger} \hat{a}_{b}^{\phantom{\dagger}}\hat{a}_{a}^{\phantom{\dagger}})}\cdots\right)^N|\psi_0\rangle.
\end{align}
We will show below, that for typical molecules one usually needs an $N$ value that is on the order of $10-20$ for an accurate representation of the operator. But, the case with $N=1$ often can produce nearly as accurate results, because the variational principle has additional freedom in it that allows it to correct some of the Trotter errors, by modifying the precise value of the amplitudes. Note that the order of the factors in the products in the parenthesis does not matter if we take the limit $N\to\infty$, but it is common to pick a particular ordering scheme, especially when working with finite values of $N$ (where the ordering does matter).

There is an exact operator identity for each of the individual UCC factors that appear in the Trotter product formula~\cite{luogen,evangelista,jia}. It arises because the operators in the exponent of a single UCC factor obey a hidden SU(2) algebra. It is
\begin{align}\label{operator_identity}
& \exp[\theta_{i_1\cdots i_n}^{a_1\cdots a_n}(\hat{a}_{a_1}^\dagger\cdots a_{a_n}^\dagger \hat{a}_{i_1}^{\phantom{\dagger}}\cdots\hat{a}_{i_n}^{\phantom{\dagger}}-\hat{a}_{i_n}^\dagger\cdots a_{i_1}^\dagger \hat{a}_{a_n}^{\phantom{\dagger}}\cdots\hat{a}_{a_1}^{\phantom{\dagger}})]\nonumber\\
&= 1 + \sin\theta_{i_1\cdots i_n}^{a_1\cdots a_n} (\hat{a}_{a_1}^\dagger\cdots a_{a_n}^\dagger \hat{a}_{i_1}^{\phantom{\dagger}}\cdots\hat{a}_{i_n}^{\phantom{\dagger}}-\hat{a}_{i_n}^\dagger\cdots a_{i_1}^\dagger \hat{a}_{a_n}^{\phantom{\dagger}}\cdots\hat{a}_{a_1}^{\phantom{\dagger}}) \nonumber\\
&+(\cos\theta_{i_1\cdots i_n}^{a_1\cdots a_n}-1)[\hat{n}_{a_1}\dots\hat{n}_{a_n}(1-\hat{n}_{i_1})\dots(1-\hat{n}_{i_n})\nonumber\\
&\quad\quad\quad\quad\quad\quad\quad\quad+(1-\hat{n}_{a_1})\dots(1-\hat{n}_{a_n})\hat{n}_{i_1}\dots\hat{n}_{i_n}]
\end{align}
for the general order-$n$ UCC factor.

The variational ansatz with $N=1$ is called the factorized form of the UCC (sometimes the factorized form of the UCC also allows individual factors to repeat, but we do not do that in this work). It is a different ansatz than the original UCC ansatz. Indeed, it now has a dependence on the ordering of the factors (because some factors do not commute with other factors). But, if the factors are chosen with a reasonable ordering scheme, then the variational principle helps make different orderings produce similar accuracies for the final energies that are calculated. But note that a specific ordering does produce constraints on the amplitudes. They no longer can be freely modified, because the de-excitations that arise as more and more factors are applied, produce constraints on the relative values of different amplitudes. For example, a particular ordering may not allow two amplitudes to be exactly the same---one amplitude may be constrained to be equal to the other plus $\sin^2\theta$---if $\theta\ne 0$, they cannot be identical.

In this work, we focus on the factorized form of the UCC and how it can be used in creating different variational wavefunction ans\"atze for electronic structure calculations. We have already seen that the Trotter product formula allows us to express the original UCC operator in terms of products of UCC factors, with factors being repeated. In this work, we explore two additional themes---the first is showing how one can perform configuration-interaction calculations on a quantum computer instead of UCC calculations. Since it is widely believed that UCC calculations will be more accurate than a CI calculation, this is really an academic exercise. But, there may be some situations where the manipulations we discuss do become important in variational state preparation and it does illustrate the flexibility one has within the factorized form of the UCC. The second is examining the accuracy of the factorized form of the UCC versus the Trotter product formula when we perform a variational minimization of the energy. This result tells us what is the most efficient ansatz to use when performing a VQE calculation on a quantum computer.

The remainder of the paper is as follows: In Sec. II, we describe how one can perform a configuration interaction calculation on a quantum computer. In Sec. III, we compare the $N=1$ form of the UCC ansatz to the exact formula for $N\to\infty$. We conclude in Sec. IV.

\section{The configuration-interaction approximation on a quantum computer}

The configuration interaction (CI) approximation works with a truncated Hamiltonian that is projected onto a specific set of determinants. Within this restricted subspace, the Hamiltonian is then diagonalized, producing a variational approximation to the true ground-state energy, and a good approximation to the ground-state, projected onto the determinants that are used in the CI basis set. The CI approximation is not generally used, except in tailored basis sets, such as the selective-CI approximation. This is because one can usually achieve higher accuracy with a CC calculation that employs the same number of amplitudes as the number of determinants in the CI. In addition, the CC approximation is size-consistent, while the CI usually is not.

Since most operators applied on a quantum computer are unitary, it seems like one cannot easily create a CI wavefunction to use in a variational calculation, but it is indeed possible to do this using the factorized form of the UCC. Each application of a UCC factor adds a determinant to the wavefunction when it acts on the reference state. It can add additional determinants when it acts on other states in the current expansion of the wavefunction. To create the CI state, we need to prune the wavefunction and remove the added determinants that are unwanted. This can be achieved via a variant of the elimination algorithm, by removing the extra terms, one-by-one.

It is best to start with a simple example, before moving to the general case. The simplest case that has this behavior is a a Hubbard model with nearest-neighbor hopping ($-t$) on a four-site ring. There are eight spin-orbitals, composed from the four single-particle eigenstates in momentum space. State 0 has $k=0$ and its energy is $-2t$, state 2 has $k=\pi$ and energy $2t$, states 1 and 3 have $k=\pi/2$ and $3\pi/2$, both with energy 0. At half filling, we choose the reference state to occupy the 0 state (both up and down) and the 1 state (both up and down), so that the reference state is $|10\bar 1 \bar 0\rangle$, where the overbars indicate the down spins.

\begin{figure}
    \centering
    \includegraphics[width=3.15in]{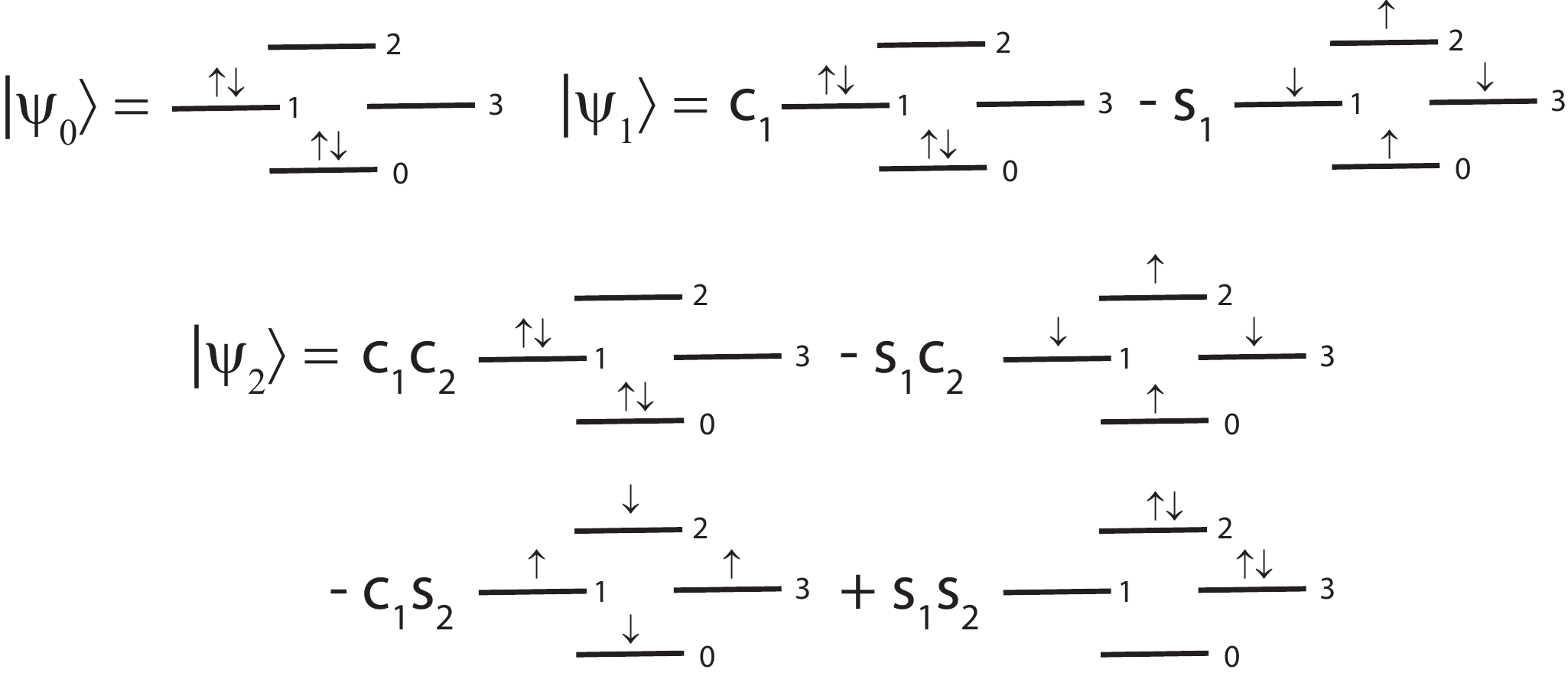}
    \caption{Schematic of the determinants created in the example discussed in the text. The lines represent the different energy levels; level 1 and level 3 are degenerate. The arrows represent the inclusion of a single-particle state in a given determinant, denoting the up-spin or down-spin state, respectively. $|\psi_0\rangle$ is the reference state and $|\psi_1\rangle$ is the state after one doubles UCC factor is applied (we use a shorthand $c_1=\cos \theta_{1\bar 0}^{2\bar{3}}$ and $s_1=\sin \theta_{1\bar 0}^{2\bar{3}}$). $|\psi_2\rangle$ is the state after applying two doubles operations, with $c_2$ and $s_2$ the corresponding trigonometric functions with argument $\theta_{3\bar 2}^{0\bar{1}}$. The state $|\psi_3\rangle$ has a similar form as $|\psi_2\rangle$, but with different coefficients (not shown).}
    \label{fig:CI}
\end{figure}

Our example is illustrated schematically in Fig.~\ref{fig:CI}.
We start with the reference state $|10\bar 1\bar 0\rangle$. We want to add the determinant $|20\bar 3\bar 1\rangle$ to the CI calculation. We do so by applying the corresponding doubles operator, to find
\begin{align}
    |\psi_1\rangle&=e^{\theta_{1\bar 0}^{2\bar{3}}\left (\hat{a}_{2\uparrow}^\dagger\hat{a}_{3\downarrow}^\dagger\hat{a}_{0\downarrow}^{\phantom{\dagger}}\hat{a}_{1\uparrow}^{\phantom{\dagger}}-\hat{a}_{1\uparrow}^\dagger\hat{a}_{0\downarrow}^\dagger\hat{a}_{3\downarrow}^{\phantom{\dagger}}\hat{a}_{2\uparrow}^{\phantom{\dagger}}\right )}|10\bar 1\bar 0\rangle\nonumber\\
    &=\cos \theta_{1\bar 0}^{2\bar{3}}|10\bar 1\bar 0\rangle-\sin \theta_{1\bar 0}^{2\bar{3}}|20\bar 3\bar 1\rangle
\end{align}
after using the exact operator identity.
By adjusting $\theta_{1\bar 0}^{2\bar{3}}$, we can have arbitrary weight for the two terms in the superposition. Note that the overall sign of the second term is determined by the ordering convention of the fermionic raising operators acting on the vacuum, that are used in determining the determinant.

Now we want to add the determinant $|31\bar 2\bar 0\rangle$.
To do this, we apply a second doubles operator to $|\psi_1\rangle$. The new state becomes
\begin{align}
     |\psi_2\rangle &=e^{\theta_{0\bar 1}^{3\bar{2}}\left (\hat{a}_{3\uparrow}^\dagger\hat{a}_{2\downarrow}^\dagger\hat{a}_{1\downarrow}^{\phantom{\dagger}}\hat{a}_{0\uparrow}^{\phantom{\dagger}}-\hat{a}_{0\uparrow}^\dagger\hat{a}_{1\downarrow}^\dagger\hat{a}_{2\downarrow}^{\phantom{\dagger}}\hat{a}_{3\uparrow}^{\phantom{\dagger}}\right )}|\psi_1\rangle  \nonumber\\
     & =\cos \theta_{1\bar 0}^{2\bar{3}}\cos \theta_{0\bar 1}^{3\bar{2}}|10\bar 1\bar 0\rangle-\cos \theta_{1\bar 0}^{2\bar{3}}\sin \theta_{0\bar 1}^{3\bar{2}}|31\bar 2\bar 0\rangle\nonumber\\
     &-\sin \theta_{1\bar 0}^{2\bar{3}}\cos \theta_{0\bar 1}^{3\bar{2}}|20\bar 3\bar 1\rangle+\sin \theta_{1\bar 0}^{2\bar{3}}\sin \theta_{0\bar 1}^{3\bar{2}}|32\bar 3\bar 2\rangle.
\end{align}
The first state is the reference, and the next two states are the two determinants we are adding into the CI calculation. But we have the fourth term, which is an extra determinant, that we did not want. So we need to remove it. One might ask why? The issue is that this extra determinant does not have a free amplitude that we can adjust. Instead, it has an amplitude determined by the amplitudes of the other two determinants that we added. This is not the way a CI calculation works, where each added determinant has its own adjustable amplitude in the superposition. It can be removed by applying a quad operator, which acts only on the first and last terms in $|\psi_2\rangle$. We find that
\begin{align}
    &|\psi_3\rangle=e^{\theta_{01\bar 0 \bar 1}^{23\bar 2 \bar 3}\left (\hat{a}_{2\uparrow}^\dagger \hat{a}_{3\uparrow}^\dagger\hat{a}_{2\downarrow}^\dagger\hat{a}_{3\downarrow}^\dagger\hat{a}_{1\downarrow}^{\phantom{\dagger}}\hat{a}_{0\downarrow}^{\phantom{\dagger}}\hat{a}_{1\uparrow}^{\phantom{\dagger}}\hat{a}_{0\uparrow}^{\phantom{\dagger}}-\hat{a}_{0\uparrow}^\dagger \hat{a}_{1\uparrow}^\dagger\hat{a}_{0\downarrow}^\dagger\hat{a}_{1\downarrow}^\dagger\hat{a}_{3\downarrow}^{\phantom{\dagger}}\hat{a}_{2\downarrow}^{\phantom{\dagger}}\hat{a}_{3\uparrow}^{\phantom{\dagger}}\hat{a}_{2\downarrow}^{\phantom{\dagger}}\right )}|\psi_2\rangle\nonumber\\
    &=\left (\cos \theta_{1\bar 0}^{2\bar{3}}\cos \theta_{0\bar 1}^{3\bar{2}}\cos \theta_{01\bar 0 \bar 1}^{23\bar 2 \bar 3}-\sin \theta_{1\bar 0}^{2\bar{3}}\sin \theta_{0\bar 1}^{3\bar{2}}\sin \theta_{01\bar 0 \bar 1}^{23\bar 2 \bar 3}\right )|10\bar 1\bar 0\rangle\nonumber\\
    &-\cos \theta_{1\bar 0}^{2\bar{3}}\sin \theta_{0\bar 1}^{3\bar{2}}|31\bar 2\bar 0\rangle-\sin \theta_{1\bar 0}^{2\bar{3}}\cos \theta_{0\bar 1}^{3\bar{2}}|20\bar 3\bar 1\rangle\nonumber\\
    &+\left (\cos \theta_{1\bar 0}^{2\bar{3}}\cos \theta_{0\bar 1}^{3\bar{2}}\sin \theta_{01\bar 0 \bar 1}^{23\bar 2 \bar 3}+\sin \theta_{1\bar 0}^{2\bar{3}}\sin \theta_{0\bar 1}^{3\bar{2}}\cos \theta_{01\bar 0 \bar 1}^{23\bar 2 \bar 3}\right )|32\bar 3\bar 2\rangle.
\end{align}
We can remove the unwanted term by choosing 
\begin{equation}
    \tan \theta_{01\bar 0 \bar 1}^{23\bar 2 \bar 3}=-\tan \theta_{1\bar 0}^{2\bar{3}} \tan \theta_{0\bar 1}^{3\bar{2}}.
\end{equation}
One can verify that the state $|\psi_3\rangle$ is normalized, and by choosing $\theta_{1\bar 0}^{2\bar{3}}$ and $\theta_{0\bar 1}^{3\bar{2}}$, we have all possible linear superpositions possible of the three determinants in the wavefunction. This is exactly what is needed for a CI calculation.

Now, suppose we have $n$ doubles determinants already in the CI approximation. To add a new doubles determinant, we use the corresponding doubles UCC factor. When this operator acts on the reference state, it creates the desired doubles determinant that is being added. The de-excitation term in the UCC factor cannot de-excite any term, because it is a doubles de-excitation and all of the other $n$ doubles terms in the superposition are \textit{different} doubles determinants. But the excitation part will excite every term in the superposition for which an excitation is allowed. This creates some number of quad excitations. We need to remove all of them to have a CI approximation. Each quad that was added, can be removed, one-by-one, by applying a similar quad UCC factor, with the amplitude chosen to ensure the coefficient of the given quad is zero. Each quad that is applied in this removal procedure can create a sextuplet excitation, when applied on every double excitation that can still be further excited. As we continue to apply additional quads UCC factors to remove the unwanted quad determinants, we will create additional sextuplet excitations, but we can also de-excite some of the previously excited sextuplets down to doubles. These doubles are always ones that already appeared in the superposition---but their coefficient is modified when this happens. Eventually, we have removed all of the offending quads. We now have a number of offending sextuplets and all of the desired doubles. We continue in the same hierarchical fashion to remove all sextuplets. This requires using a sextuplet UCC factor. Again, all possible doubles that can be excited to octuplets will be so excited. Removing additional sextuplets can modify the coefficients of the doubles again. And the procedure continues. Will it ever stop? Yes, it must. This follows either from the elimination algorithm~\cite{evangelista}, or from the simple fact, that because we use a finite basis set of allowed orbitals, there is a maximal excited determinant that we can have (we cannot excite to an order higher than the number of electrons in the original reference state).

So, this approach will eventually remove all higher order determinants, leaving behind only the desired doubles determinants. The only remaining question, for this to be an unbiased CI approximation, is whether the coefficients of the different doubles determinants have independent amplitudes that can be freely varied. While this should be true, it is a subtle point, that does not have any simple answer without calculating the different coefficients concretely and seeing if there are any extraneous constraints on them (similar to what we did with the example above). This appears to be unlikely, but we cannot rule it out at this stage. However, if this does appear to cause a problem, one should be able to adjust the doubles coefficient by applying a correction UCC doubles factor for the problematic coefficient. This will require additional quads and higher-order corrections to finally reduce to having just doubles again. 

Suppose we have added all desired doubles and now we wish to add in other determinants, such as singles. The first single added, will also add in a number of triples because it can excite many of the doubles. These triples can be removed following a similar strategy as described above. If we add two singles, then we will have extra doubles excitations in addition to the extra triples. Again, following a hierarchical elimination procedure, we can remove all extraneous terms. Next, if we add triples terms, they can created quads and quintuplets. These can also be removed as before.

One can see that the procedure becomes quite complicated as more and more determinants are added into the CI calculation. In addition, it is possible that some of the determinants kept in the CI calculation may not have completely independent coefficients. While the elimination algorithm suggests that these can ultimately be made completely independent of each other, it is not clear precisely how to guarantee this. So, modulo some possible constraints on the coefficients of the desired terms in the CI approximation, one can perform CI-based calculations on a quantum computer. We do note that the complexity of carrying this out, and the fact that a coupled-cluster calculation is likely to be more accurate than a corresponding CI calculation, means it is unlikely that this would widely used. However, there may be some specialized situations where it proves to be valuable. We find it  interesting that the factorized form of the UCC allows us to work with such an approximation.

\section{The Trotter approximation to the conventional Unitary coupled cluster approximation}

The conventional UCC approximation, where we apply $e^{\hat{T}-\hat{T}^\dagger}$ to our reference state, is a uniquely formed wavefunction ansatz---it does not depend on the ordering of the terms in the operator $\hat{T}$. But, if we convert it to an approximate form expressed in terms of individual UCC factors, then the ordering plays a role, as we discussed above. In this section, we discuss accuracy issues associated with approximating the conventional UCC approximation with a Trotter product formula that has a finite value of $N$. Note if we wish to approximate the conventional UCC approximation with a specific approximation that has a rigorously bounded error, then we think of the Trotter product formula as being an approximation that becomes more and more accurate as $N$ is increased. In this case, we need to know how large does $N$ need to be to achieve our desired chemical accuracy?

We can look at this problem in a different way. We can think of it as we think of the quantum approximate optimization algorithm (QAOA)~\cite{qaoa}, which seeks the most accurate approximation given the number of levels (that is, the number of UCC factors) in the wavefunction ansatz. In this case, we may find a wavefunction that gives a more accurate energy than we would have if we identified the amplitudes in the factorized form with the same amplitudes that we would use in the conventional UCC ansatz. This is because, by varying the values of the amplitudes, we can sometimes correct issues associated with Trotter product formula errors. The best way to investigate this is by looking at a concrete example. Of course, this is a case study and not a rigorous proof for the general situation.

The problem we choose to test these ideas on is one that can be solved by a full CI calculation. We choose a particularly small system in order to be able to perform all calculations exactly and efficiently.  We look at the open H$_6$ chain. We use the STO-6G basis set. The system is a modest size, with 12 spin orbitals and a Hilbert-space dimension of 400.  Our exact ansatz includes all possible excitations allowed by the number of electrons and the total number of orbitals in the operator $\hat{T}$. We choose the interatomic spacing to be $4\text{\AA}$ to be in the strong-correlation regime.

We evaluate the Trotter product formula in two different ways. The first way, chooses the ordering within each Trotter factor to be ordered in terms of the most important amplitudes, as determined by an MP2 calculation (for the singles and doubles) and as determined by the energy of the excitation for all triples, then all quads, and so on. This ordering is then repeated $N$ times to obtain the Trotter product formula. The other way we do it is to pick the UCC factors at random for one Trotter step and then repeat the same ordering for the remaining $N$ Trotter steps. This is motivated by work on the Trotter product formula in time evolution, which showed that picking Trotter factors at random (and using importance sampling) improved the accuracy of the Trotter product formula for a fixed number of Trotter factors~\cite{random}. We will see that does not occur here. Finally, we perform a full optimization using the $N=1$ Trotter product formula, including only singles and doubles excitations and de-excitations; the doubles amplitudes are chosen in the MP2 order, followed by the singles amplitudes. 

The way that we choose the exact amplitudes for the conventional UCC ansatz, is to start from the exact ground state, as determined by an FCI approximation. Then, because we can calculate the conventional UCC ansatz exactly, we fit the amplitudes to give us the exact ground state for $e^{\hat{T}-\hat{T}^\dagger}|\psi_0\rangle$. This is done by first computing the matrix of $\hat{T}-\hat{T}^\dagger$ in the given product-state basis. The exponential of the matrix was then calculated with the  SciPy\cite{2020SciPy-NMeth} package. Note that since we have fixed the amplitudes, there is no optimization performed during these calculations. Just an evaluation of the operators acting on the reference state and then calculating the energy expectation value.

\begin{figure}
    \centering
    \includegraphics[width=2.75in]{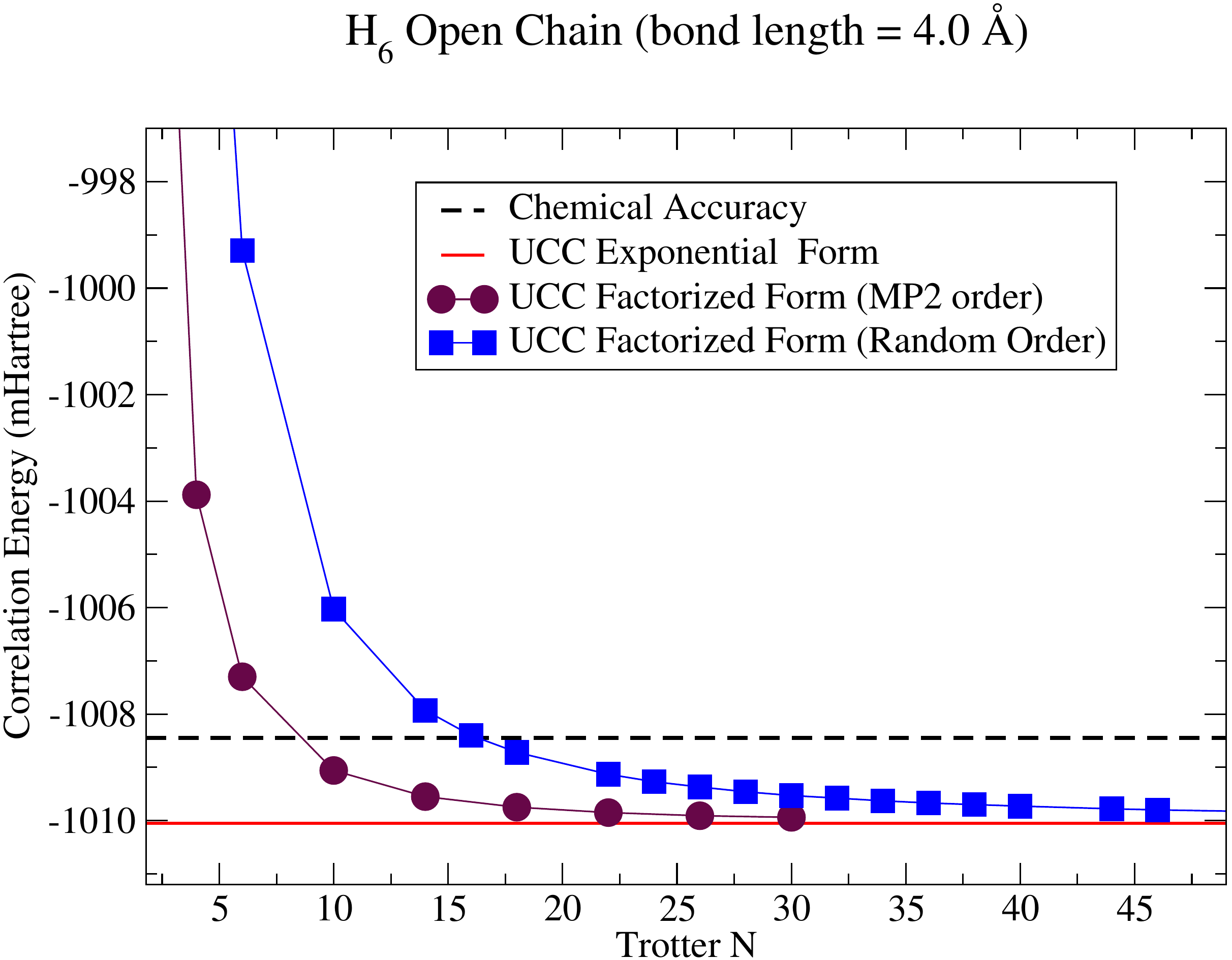}
    \caption{Comparison of the Trotter product formula to the exact energy of an H$_6$ chain with an interatomic spacing of $4\text{\AA}$. The two curves represent the cases where the UCC factors are chosen in an ordering according to the MP2 perturbation theory (purple) or randomly (blue). Chemical accuracy is indicated by the dashed line.
    The full CI result (and the full UCC ansatz, which becomes exact) is given by the red line.}
    \label{fig:trotter}
\end{figure}

In Fig.~\ref{fig:trotter}, we show the results for the accuracy of the correlation energy as a function of the number of Trotter steps $N$. One can see that for this simple problem, we require $N$ to be on the order of 10 to achieve chemical accuracy. Interestingly, a variationally optimized UCC calculation in the factorized form with  $N=1$ and including only singles and doubles (in MP2 order), produces a correlation energy of $-1003.082$~mH, where we optimize the amplitudes to produce the best energy. This is to be compared with the FCI correlation energy of $-1010.085$~mH. This shows that the $N=1$ approximation is quite accurate (but not quite chemical accuracy), even though it does not represent a good approximation of the conventional UCCSD ansatz! For this stretch, the conventional CC ansatz is not accurate. We discuss the boost in accuracy due to the variational principle next.

\begin{figure}
    \centering
    \includegraphics[width=2.75in]{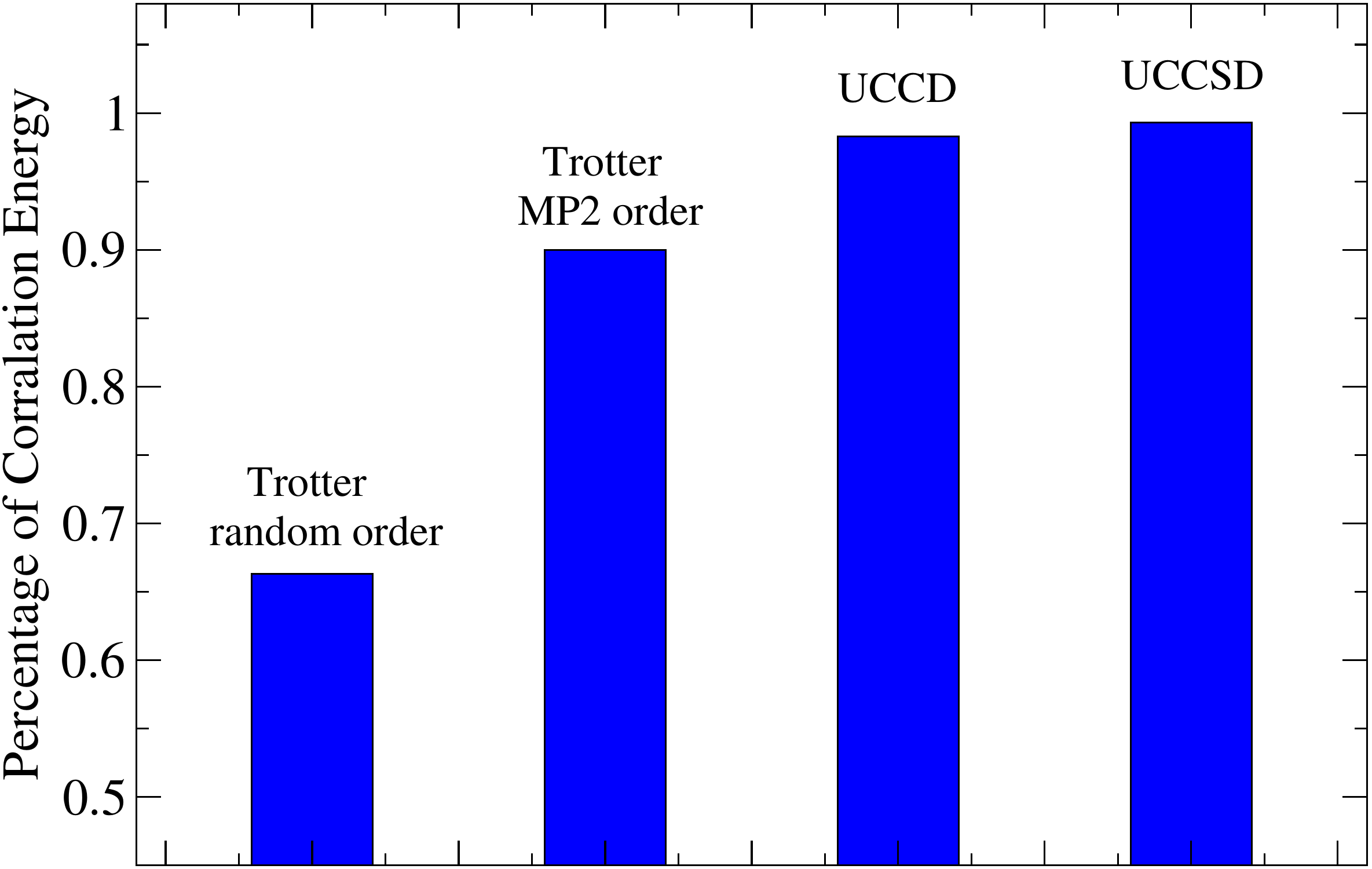}
    \caption{Percentage of the correlation energy for the different approximations for a single-step Trotter formula ($N=1$); these results are not optimized, since the amplitudes are known exactly from the exact ground state. The UCC(S)D results are the optimized results for the $N=1$ ansatz with just doubles or with just singles and doubles.}
    \label{fig:corr-energy}
\end{figure}

In Fig.~\ref{fig:corr-energy}, we show the percentage of the correlation energy that is found for the different $N=1$ approximations. The worst result comes from the $N=1$ approximation to the conventional UCC ansatz when we choose the UCC factors in random order. Choosing them in the MP2 order does significantly better, indicating that the ordering of the UCC factors can play a significant role. If we do not use the exact amplitudes in the conventional UCC ansatz, but instead optimize the amplitudes, we get the next two bars, corresponding to a doubles-only ansatz and a singles and doubles ansatz. One can clearly see that the variational principle allows for significant improvements on the accuracy of the total energy, when we use the factorized form of the UCC as the wavefunction ansatz.

The results shown here are suggestive that one requires moderate to large $N$ values to correctly approximate the conventional UCC ansatz. But, we can still achieve high accuracy with $N=1$ if we use the UCC ansatz in its factorized form and perform an optimization to minimize the energy. Just like in the QAOA approach, we find the optimization step greatly improves the accuracy of the final answer. It does this by partially compensating for the Trotter error of the $N=1$ form of the ansatz.

\section{Conclusions}

In this work, we showed that the factorized form of the UCC has great flexibility as a wavefunction ansatz for deployment on quantum computers. It can produce a conventional, or selective CI wavefunction. It can produce the conventional UCC wavefunction. Or, it can produce something new, that balances ease of implementation with high accuracy, which is attained through the optimization step for the total energy. This implies that if one wants to use a wavefunction ansatz in a fermionic form, then the $N=1$ Trotter product formula, with the doubles amplitudes chosen in the MP2 order, is likely to produce high accuracy with low circuit depth. If the accuracy is not sufficient, then triples and higher-order excitations can be added in using the same factorized ansatz.
Our work suggests that this is a general principle for carrying out variational quantum eigensolver calculations on a quantum computer.

\begin{acknowledgements}
 JC and HPC are supported by the Department of Energy, Basic Energy Sciences as part of the Center for Molecular Magnetic Quantum Materials, an Energy Frontier Research Center under Award No. DE-SC0019330. JKF is supported from the National Science Foundation under grant number CHE-1836497. JKF is also funded by the McDevitt bequest at Georgetown University. This research used resources of the National Energy Research Scientific Computing Center (NERSC), a U.S. Department of Energy Office of Science User Facility operated under Contract No. DE-AC02-05CH11231, and University of Florida Research Computing systems.
 We also acknowledge useful discussions with Rodney Bartlett and Dominika Zgid.
\end{acknowledgements}

\section*{Data Availability Statement}

Data available in article or supplementary material.

\nocite{*}
\bibliography{refs}

\end{document}